\documentclass[aps,prb,twocolumn,floatfix, superscriptaddress]{revtex4-2}
\usepackage{times}
\usepackage{epsfig}
\usepackage{amsmath}
\usepackage{amssymb}
\usepackage{graphicx}
\usepackage{bbold}
\usepackage{hyperref}
\usepackage[switch]{lineno}
\usepackage{pifont}
\usepackage{bbding}
\hypersetup{colorlinks=true,citecolor=red,linkcolor=blue}

\newcommand{\tsh}{t_\mathrm{h}}
\newcommand{\hatj}{\hat{j}}

\begin{document}
\title{Site-selective doublon-holon dynamics in a pumped one-dimensional Hubbard superlattice \\
with staggered Coulomb interactions}
\author{Zhenyu Cheng}
\author{Ying Li}
\affiliation{College of Physics and Technology, Guangxi Normal University, Guilin, Guangxi 541004, China}
\author{Hantao Lu}
\affiliation{Lanzhou Center for Theoretical Physics, Key Laboratory of Theoretical Physics of Gansu Province, and Key Laboratory of Quantum Theory and Applications of MoE, Lanzhou University, Lanzhou, Gansu 730000, China}
\author{Xiang Hu}
\affiliation{College of Physics and Technology, Guangxi Normal University, Guilin, Guangxi 541004, China}
\author{Zhongbing Huang}
\affiliation{Department of Physics, Hubei University, Wuhan 430062, China}
\author{Gregory A. Fiete}
\affiliation{Department of Physics, Northeastern University, Boston, Massachusetts 02115, USA}
\affiliation{Department of Physics, Massachusetts Institute of Technology, Cambridge, Massachusetts 02139, USA}
\author{Liang Du}
\email{Liang Du: liangdu@gxnu.edu.cn}
\affiliation{College of Physics and Technology, Guangxi Normal University, Guilin, Guangxi 541004, China}
\begin{abstract}
Doublon-holon dynamics is investigated in a pumped one-dimensional Hubbard model with a staggered on-site Coulomb interaction at half-filling.
When the system parameters are set to be in the Mott insulating regime the equilibrium sublattice density of states exhibits several characteristic peaks, corresponding to the lower and upper Hubbard bands as well as hybridization bands. We study the linear absorption spectrum and find two main peaks characterizing the photon frequencies which excite the ground state to an excited state.
For a system driven by a laser pulse with general intensity and frequency, both the energy absorption and the doublon-holon dynamics exhibit distinct behaviors as a function of laser amplitude and frequency. 
Single-photon processes are observed at low laser intensity where the energy is absorbed for resonance laser frequencies. For strong laser intensity multi-photon induced dynamics are observed in the system, which are confirmed by an evaluation of the Loschmidt amplitude. 
The contribution of multi-photon processes to site-specific double occupancy is also characterized by the generalized Loschmidt amplitude. The site-selective doublon-holon dynamics are observed in both the one and multi-photon processes and the site-selective behavior is explained within a quasiparticle picture.
Our study suggests strategies to optically engineer the doublon-holon dynamics in one dimensional strongly correlated many-body systems. 
\end{abstract}
\date{\today}
\maketitle
\def\thefootnote{*}\footnotetext{These authors contributed equally to this work}\def\thefootnote{\arabic{footnote}}
\section{INTRODUCTION}
\label{sec:intro}
Non-equilibrium control of quantum states in strongly correlated systems with optical techniques and its physical understanding have been attracting attention from the condensed matter community in the past decade and remains challenging\cite{Aoki:rmp2014, Giannetti:ap2016, Heyl:rpp2018}.
Of particular research interests are: (1) thermalization and prethermalization behavior as the system is driven far from equilibrium\cite{Aoki:rmp2014,Mori:jpb2018};
(2) non-thermal states not accessible in equilibrium\cite{Berges:prl2008,Chiocchetta:prl2017,Heyl:rpp2018};
(3) non-equilibrium control of quantum phase transitions \cite{Fausti:sci2011,Giannetti:ap2016,Cavalleri:cp2018,Delatorre:rmp2021,Budden:np2021,Rowe:np2023}; and
(4) disentangling degrees of freedom in the strongly correlated electronic system\cite{Dalconte:sci2012}.

In strongly correlated systems, the dynamics of the low-dimensional (one and two spatial dimensions) Hubbard model has been studied extensively, with a number of exotic behaviors observed. 
For example, in the pumped one-dimensional Hubbard model, unconventional superconductivity\cite{Kaneko:prl2019}, a photo-induced insulator to metal transition\cite{Ejima:prr2022}, and charge, spin and $\eta-$spin separation\cite{Murakami:prl2023} have been revealed.  
Furthermore, in laser-pumped systems the evolution of doubly occupied states (doublons) are dependent on the laser frequency\cite{Innerberger:epjp2020}. Such a non-equilibrium system can be described as a generalized Gibbs ensemble\cite{Murakami:cp2022}.
On the other hand, in a two-dimensional Hubbard cluster, the system exhibits Rabi-like oscillations\cite{Okamoto:sr2021}, where the oscillation frequency increases with the drive amplitude (intensity).
Recently, d-wave superconductivity has been observed in a driven two-dimensional plaquette Hubbard model\cite{zhangY:prb2023}. 
In contrast to the one-dimensional case, the doublon dynamics in the Hubbard two-leg ladder exhibits weak coupling with magnetic excitations\cite{Diasdasilva:prb2012}. In addition, impact ionization is observed in the two-dimensional cluster\cite{Kauch:prb2020,Maislinger:prb2022,Gazzaneo:prb2022,Watzenbock:prb2022} where the double occupancy rises further after the laser pulse.

The non-equilibrium  studies summarized above are focused on systems with spatial homogeneity in the Hamiltonian. 
In such systems, doublons and holons are generated when a pump light is applied, which leads to an increase in the total energy of the system. After the passing of the pump light through the system, the system begins to relax through the recombination of doublons and holons, where prethermalization behavior occurs\cite{Murakami:cp2022}. 
By contrast, the non-equilibrium dynamics in systems without spacial homogeneity  (e.g., ionic Hubbard model, Hubbard superlattice with a staggered Coulomb interaction) have received relatively little attention. 
For the purpose of simplifying the physical picture and the analysis in nonequilibrium studies of doulon-holon dynamics, we focus our attention on the Hubbard superlattice with a staggered Coulomb potential rather than the ionic Hubbard model where charge density wave and spontaneous dimerized state\cite{Fabrizio:prl1999} could occur and complicate the analysis.

\begin{figure}[ht]
\centering
\includegraphics[angle=-0,width=0.49\textwidth]{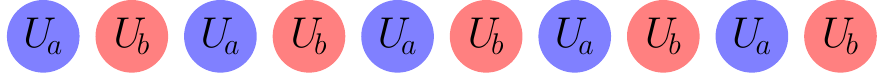}
\caption{(Color online) An example of a lattice with 10 sites, where the onsite Coulomb interaction strength is modulated alternatively (staggered) with $U_a\neq U_b$ for different sublattices. }
\label{Fig0}
\end{figure}

In this paper we focus on the non-equilibrium dynamics of the Hubbard superlattice with spatially staggered Coulomb interactions, as shown in Fig.\ref{Fig0}.
In equilibrium such a Hubbard superlattice can be experimentally realized in condensed matter with nano-scale spatial inhomogeneity\cite{Paiva:prl1996,Paiva:prb1998,Paiva:prb2000,Paiva:prb2002,Duan:jpcm2010,ZhangLL:cpb2015}.
The simplest spatially inhomogeneous case is where there are two sites in the unit cell with on-site Coulomb interactions $U_a$ for the A-sublattice and $U_b$ for the B-sublattice.
For a non-interacting B sublattice,  $U_b = 0$, and finite $U_a$ in one-dimension, the system exhibits a correlated-metallic phase at the particle–hole symmetrical point with zero spin and zero charge gaps\cite{Duan:jpcm2010,ZhangLL:cpb2015}.
For a B-sublattice with finite Coulomb interaction strength  $U_b > 0$, the system is an antiferromagnetic Mott insulator whis with zero spin gap and finite charge gap\cite{Duan:jpcm2010}.
Besides realizations in condensed matter systems, a spatial modulation of the interaction has also been reported in ${}^{174}$Yb gas systems\cite{Yamazaki:prl2010}, and the system undergoes a Mott metal-insulator transition as the Coulomb interaction increases\cite{Saitou:jsnm2013,Koga:jpsj2013,Hoang:jpsj2016} in the infinite dimensional Bethe lattice, as shown within the framework of dynamical mean-field theory\cite{Georges:rmp1996}.

In this work, by introducing a spatial inhomogenity with spatially alternating (staggered) on-site Coulomb interactions, we study the doublon-holon dynamics as a function of laser intensity and frequency in the one-dimensional Hubbard superlattice with $U_b =18.0t_h, U_b=3.0t_h$ (see Eq.\eqref{eq:eqH}), where the system is a Mott insulator before the pump is applied. 
For weak laser intensity where linear response theory applies, we observe a site-selective doublon-holon dynamics at laser frequency $\hbar\Omega \approx 3.2 t_h$, where the double occupancy of the B-sublattice is enhanced by the laser pulse, while the double occupancy of A-sublattice remains almost unchanged during and after the laser pulse. 
With strong laser intensity, multi-photon effects are observed, and the site-selected doublon enhanced for laser frequency at $\hbar\Omega = 9.4t_h$, with site-A enhanced substantially while site-B remains unchanged, in contrast to the selective behavior observed at $\hbar \Omega=3.2t_h$.  This observation opens the possibility through appropriate laser protocol to optically engineer doublon-holon states in many-body interacting systems.

Our paper is organized as follows.
In Sec.\ref{sec:model}, we describe the Hamiltonian of the pumped one-dimensional Hubbard model with modulated (staggered) site-dependent Coulomb interaction and the time-dependent Lanczos method of solution.
In Sec.\ref{sec:eqlabs}, the equilibrium sublattice site-specific density of states and the linear absorption spectrum 
are obtained using exact diagonalization.
In Sec.\ref{sec:neqdynamics}, we study the non-equilibrium dynamics of doublon-holons as a function of laser frequency and amplitude, and we study the eigenstate spectrum using the Loschmidt amplitude.
In Sec.\ref{sec:gla}, the generalized Loschmidt amplitude is introduced to exhibit details of the site-specific doublon-holon dynamics.
Finally, in Sec.\ref{sec:concl} we present the main conclusions of the paper.

\section{model and method}
\label{sec:model}
In equilibrium the one dimensional Hubbard model on the superlattice (A-B sublattice) is written as \cite{Duan:jpcm2010,ZhangLL:cpb2015},
\begin{align}
    H &= -\tsh \sum_{i\sigma}\left(c_{i\sigma}^\dagger c_{i+1\sigma}^{} + h.c.\right) + \frac{\epsilon}{2}\sum_{i,\sigma}(n_{2i,\sigma} - n_{2i+1,\sigma})\nonumber\\
    &+ U_a \sum_i n_{2i+1,\uparrow}n_{2i+1\downarrow} +  U_b \sum_i n_{2i,\uparrow}n_{2i\downarrow},
\label{eqH}
\end{align}
where $c_{i\sigma}^\dagger$ ($c_{i\sigma}$) create (annihilate) a fermionic particle at site $i$ with spin projection
$\sigma$ and $n_{i\sigma} = c_{i\sigma}^\dagger c_{i\sigma}$ for $\sigma = \uparrow, \downarrow$. Here
$\tsh$ is the anisotropic hopping amplitude connecting nearest neighbour sites and
$U_a (U_b)$ is the onsite Coulomb interaction between $\uparrow$ and $\downarrow$ spin electrons for odd (even) sites in the one-dimensional chain.
The energy $\mp \epsilon/2$ is the onsite orbit energy for the A-B sublattice.
The Hamiltonian is particle-hole symmetric at $\epsilon = (U_a - U_b)/2$ and can be rewritten as,
\begin{align}
    H &= -\tsh \sum_{i\sigma}\left(c_{i\sigma}^\dagger c_{i+1\sigma} + h.c.\right) \nonumber\\
      &+ \sum_i U_i \left(n_{i\uparrow}-\frac{1}{2}\right)\left( n_{i\downarrow}-\frac{1}{2}\right),
\label{eq:eqH}
\end{align}
where $U_i$ is the site dependent Coulomb interaction, which we choose $U_i = U_a (U_b)$ for odd (even) sites in the one dimensional chain.

In this paper, the non-equilibrium calculations are done in the particle-hole symmetric limit. We set $\tsh =1 $ as the energy unit and correspondingly, the unit of time is then the inverse of energy, $\tsh^{-1}$. The site-specific density of states is defined as,
\begin{align}
\rho_\alpha(\omega) = \sum_{i\in\alpha,\sigma} \sum_{\phi} 
    &|\langle\phi|c_{i\sigma}^\dagger|\psi_0\rangle|^2 \delta(\omega - E_\phi + E_0)  \nonumber\\
   +& |\langle\phi|c_{i\sigma}       |\psi_0\rangle|^2 \delta(\omega + E_\phi - E_0)
   \label{eq:eqdos}
\end{align}
where $\{|\phi\rangle\}$ are the eigenstates of the equilibrium Hamiltonian in Eq.\eqref{eq:eqH} with respect to energy eigenvalues $E_\phi$, and $|\psi_0\rangle$ is the ground state with energy $E_0$. The definition of the linear absorption spectrum is \cite{Okamoto:njp2019,Kaneko:prl2019,Ejima:prr2022},
\begin{align}
    \alpha(\omega) = -\frac{1}{\pi} \mathrm{Im} \langle \psi_0| \hatj \frac{1}{\omega - (H - E_0)} \hatj |\psi_0\rangle,
    \label{eq:labs}
\end{align}
and the current density operator is defined as,
\begin{align}
    \hatj = i\tsh \sum_{i \sigma} (c_{i\sigma}^\dagger c_{i+1,\sigma}^{} - c_{i+1\sigma}^\dagger c_{i,\sigma}^{}).
\end{align}
The equation for linear absorption, Eq.\eqref{eq:labs}, only works in low laser intensity regime, where linear response theory is applicable. 

We consider a system exposed to an external laser pulse with vector potential (directed along the chain direction),
\begin{align}
    A(t) = A_0 \exp[-(t-t_p)^2/2t_d^2] \cos[\Omega (t-t_p)],
\end{align}
where $A_0$ is the laser intensity, $\Omega$ is the laser frequency and the laser pulse is peaked at $t_p$ with $t_d$ characterizing the duration time (pulse width) of light.
The time-dependent Hamiltonian is written using the Peierls substitution,
\begin{align}
    H(t) &= -\tsh \sum_{i\sigma}\left( \exp [i A(t)]c_{i\sigma}^\dagger c_{i+1\sigma} + h.c.\right) \nonumber\\
    &+ \sum_i U_i \left(n_{i\uparrow}-\frac{1}{2}\right)\left( n_{i\downarrow}-\frac{1}{2}\right).
\label{neqH}
\end{align}
We choose the chain size to be $L=10$ with periodic boundary conditions and a coarse-grained time $\delta t = 0.005 \tsh^{-1}$.
In the following, we restrict ourselves to the case of half-filling with periodic boundary conditions,
where the total number of electrons $N$ is equal to the number of sites in the chain $L$.
Furthermore, we assume the total magnetization in the system vanishes,
which means the number of up-spin electrons $N_{\uparrow}$ is equal to the down-spin electrons $N_{\downarrow}$.

The exact diagonalization method (a standard Lanczos procedure) is employed to numerically calculate the ground state of the Hamiltonian at time $t=0^-$, where the lase pulse is not yet applied to the system.
The ground state is used as an initial state for the time dependent Schr\"odinger equation $i\partial_t |\Psi(t)\rangle = H(t) |\Psi(t)\rangle$.
The time evolution is implemented step-by-step based on the time-dependent Lanczos method \cite{Park:jcp1986,Mohankumar:cpc2006,Balzer:jpcm2012,Lu:prl2012,Innerberger:epjp2020},
\begin{equation}
  |\Psi(t+\delta t)\rangle \approx e^{-i H(t)\delta t} |\Psi(t)\rangle
                           \approx \sum_{l=1}^M e^{-i \epsilon_l^{} \delta t} |\Phi_l\rangle\langle \Phi_l| \Psi(t)\rangle,\nonumber
\end{equation}
where $\epsilon_l^{}$ ($\Phi_l$) are the eigenvalues (eigenvectors) of the tri-diagonal matrix generated by Lanczos iteration with $M \leq 100$. (In general, the desired M for achieving a given accuracy depends on the setups and chosen time step sizes \cite{Moler:siamr2003}.)
We set the time step size $\delta t = 0.005t_h^{-1}$ in our calculation of the time evolution.
The physical observable are computed as,
\begin{align}
     \langle O(t)\rangle = \langle \Psi(t)| O |\Psi(t)\rangle.
\end{align}

For the case of high laser intensity, multi-photon processes appear or even dominate the optical excitation in the system.
To validate the multi-photon processes and provide more details on the double-holon dynamics, the Loschmidt amplitude is used to calculate the spectral density\cite{Kennes:prb2020,Maislinger:prb2022,Okamoto:sr2021},
\begin{equation}
     L(\omega, t) = \sum_{n}|\langle E_n|\Psi(t)\rangle|^2 \delta[\omega - (E_n - E_g)]
\end{equation}
where $|E_n\rangle$ and $E_n$ are the eigenstates and eigenenergies ($E_g$ is the many-body ground state energy) of the unperturbed Hamiltonian (after pulse Hamiltonian), and $|\Psi(t)\rangle$ is the time evolved state at time $t$. Note that, the eigenstate spectrum is kept unchanged as a function of time after the pump pulse has decayed, because $H(t) = H(0)$ and as a result $|\langle E_n |\Psi(t)\rangle|$ is unchanged. 
\begin{figure}[ht]
\centering
\includegraphics[angle=-0,width=0.45\textwidth]{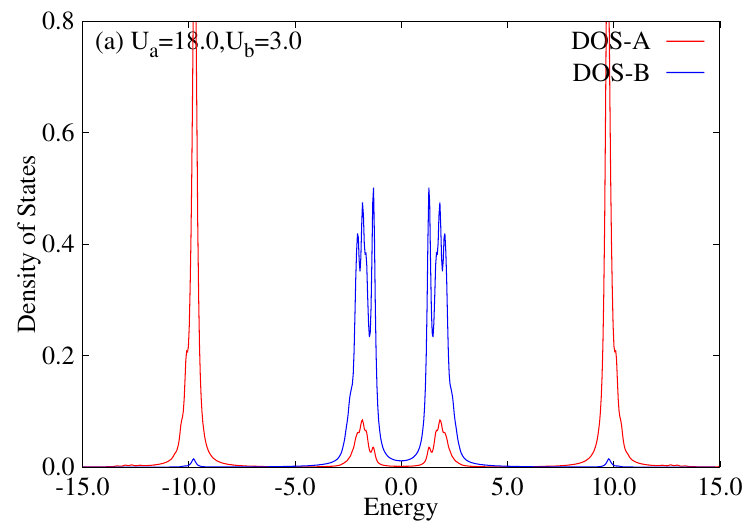}
\includegraphics[angle=-0,width=0.45\textwidth]{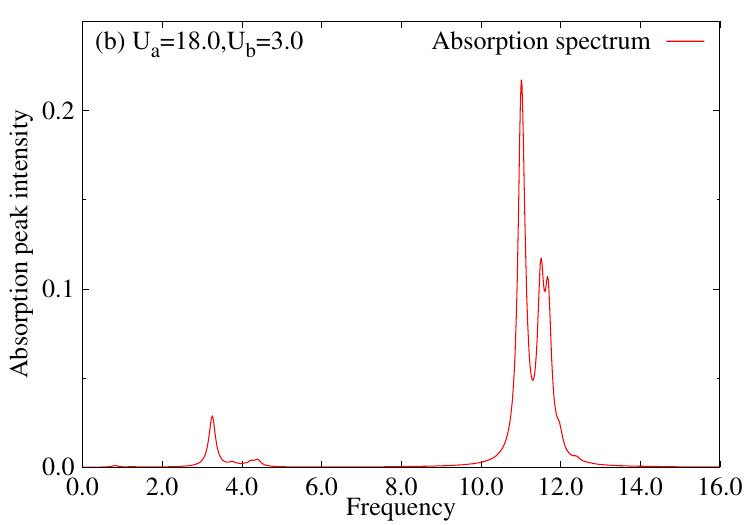}
\caption{(Color online) For the Hubbard super-lattice with fixed site-dependent Coulomb interaction strength $U_b = 3.0$ and $U_a = 18.0$, (a) The sub-lattice specified density of states in equilibrium. (b) The linear absorption spectrum as a function of frequency calculated using Eq.\eqref{eq:labs}, where linear response theory apply. }
\label{Fig:eqdos}
\end{figure}
\begin{figure}[ht]
\centering
\includegraphics[angle=-0,width=0.5\textwidth]{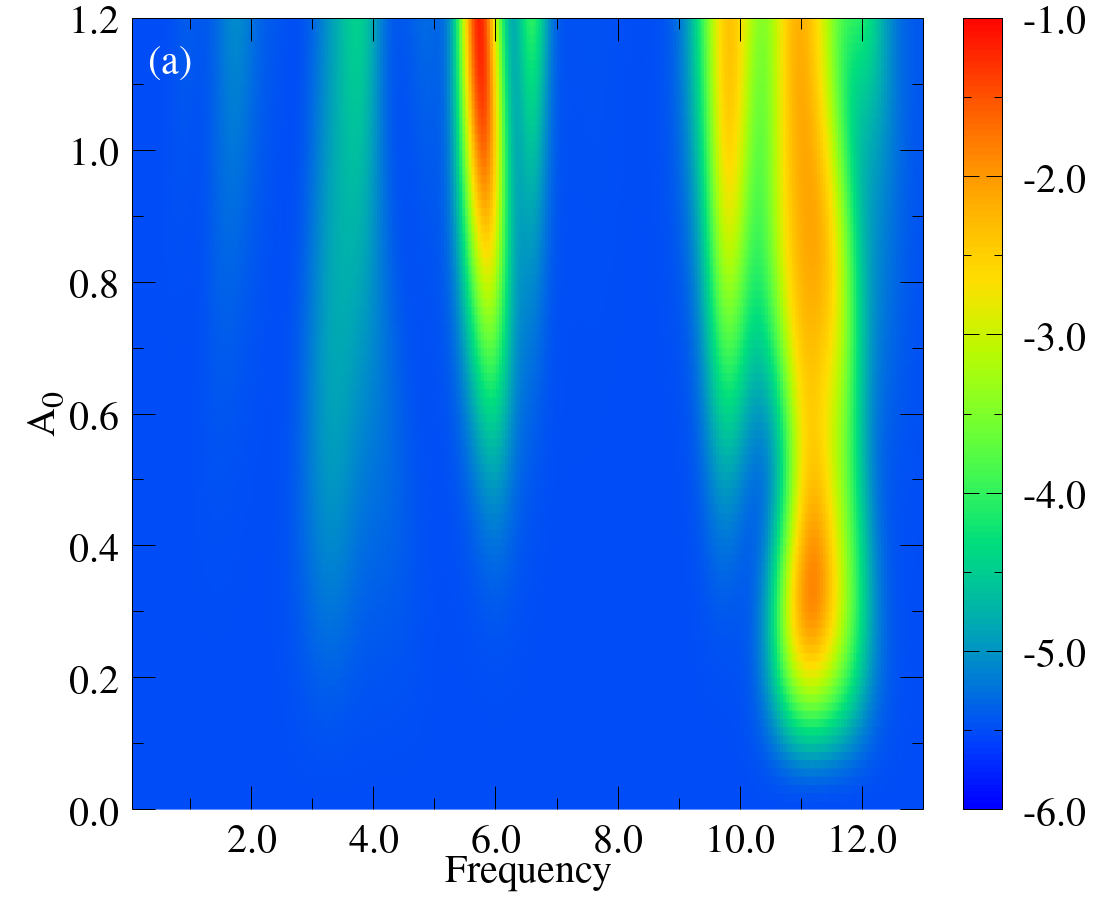}
\includegraphics[angle=-0,width=0.45\textwidth]{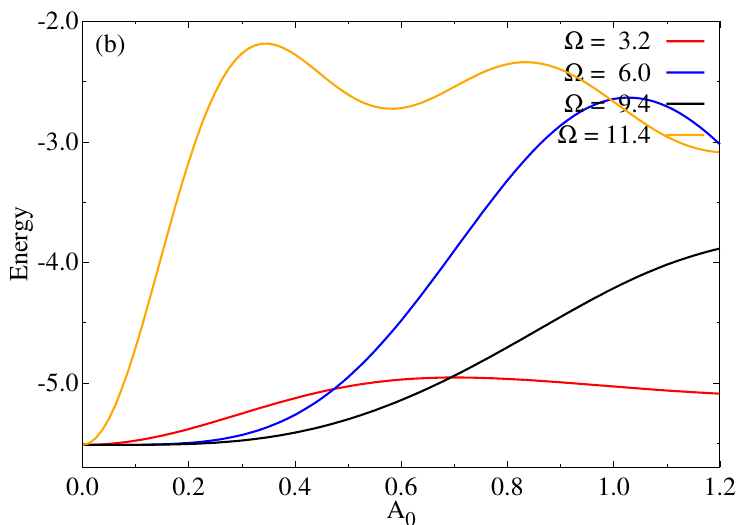}
\caption{(Color online) For the Hubbard super-lattice with fixed site-dependent Coulomb interaction strength $U_b = 3.0$ and $U_a = 18.0$,  (a) The after pulse ($t= 15.0$ used) energy as a function of laser frequency and amplitude. (b) the after pulse energy as a function of laser amplitude with fixed laser frequency $\Omega = 3.2, 6.0, 9.4, 11.4$.}
\label{Fig1}
\end{figure}
\begin{figure*}[ht]
\centering
\includegraphics[angle=-0,width=0.99\textwidth]{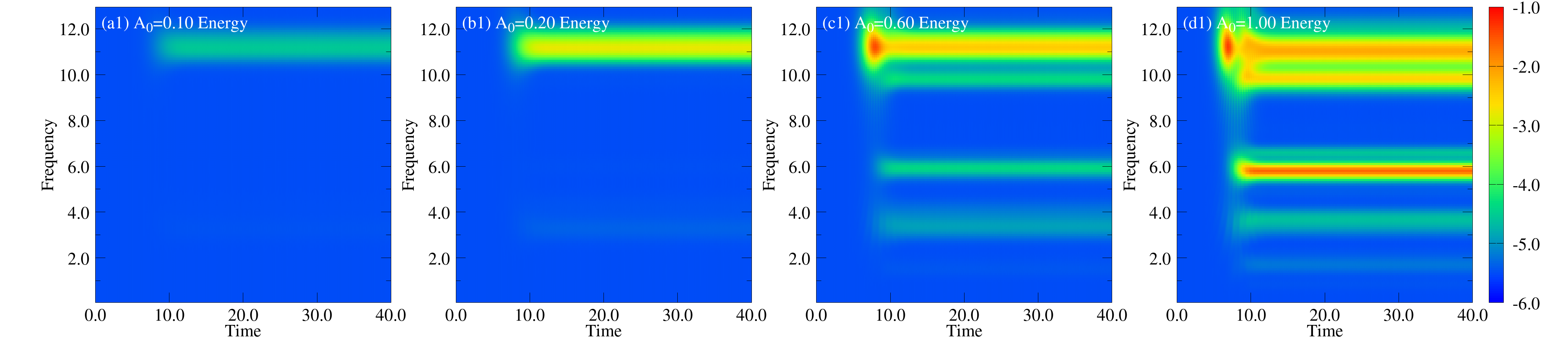}
\includegraphics[angle=-0,width=0.99\textwidth]{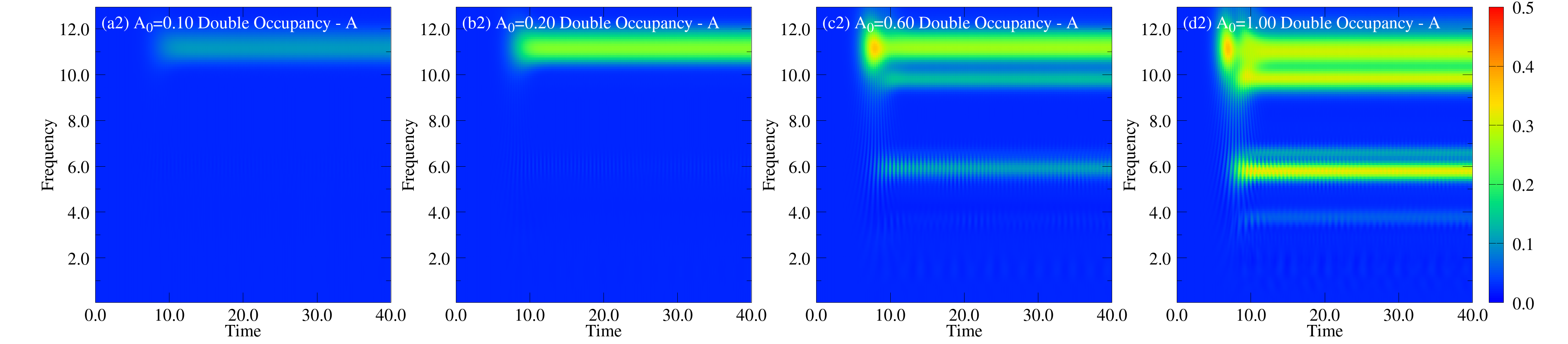}
\includegraphics[angle=-0,width=0.99\textwidth]{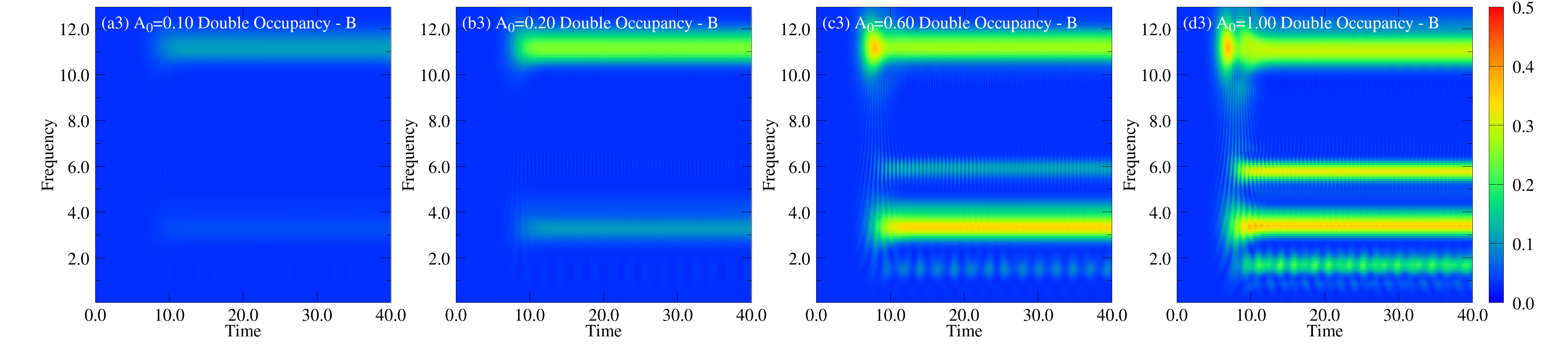}
\caption{(Color online) For the Hubbard super-lattice with fixed site-dependent Coulomb interaction strength $U_b = 3.0$ and $U_a = 18.0$, The time evolution of total energy and site-specified double occupancy as a function of frequency $\Omega$ are plotted in one column with laser intensity $A_0 = 0.1$ (a1-a3), $A_0 = 0.2$ (b1-b3), $A_0 =0.6$ (c1-c3), $A_0 =1.0$ (d1-d3).}
\label{Fig:ewt}
\end{figure*}
\begin{figure*}[ht]
\centering
\includegraphics[angle=-0,width=0.243\textwidth]{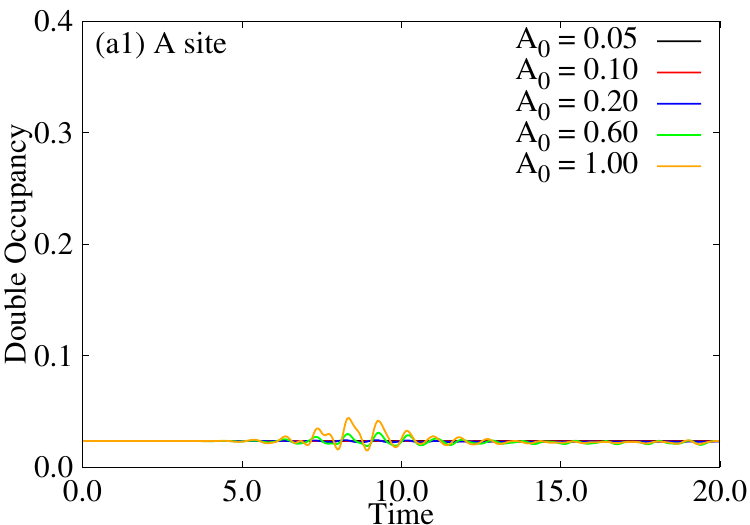}
\includegraphics[angle=-0,width=0.243\textwidth]{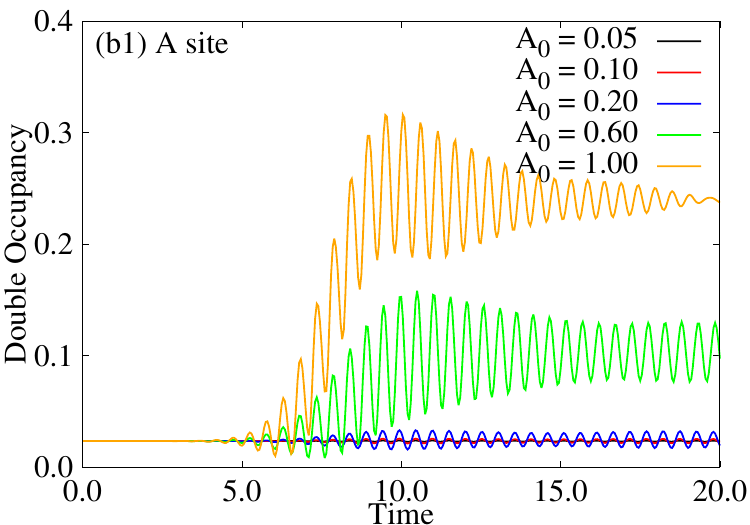}
\includegraphics[angle=-0,width=0.243\textwidth]{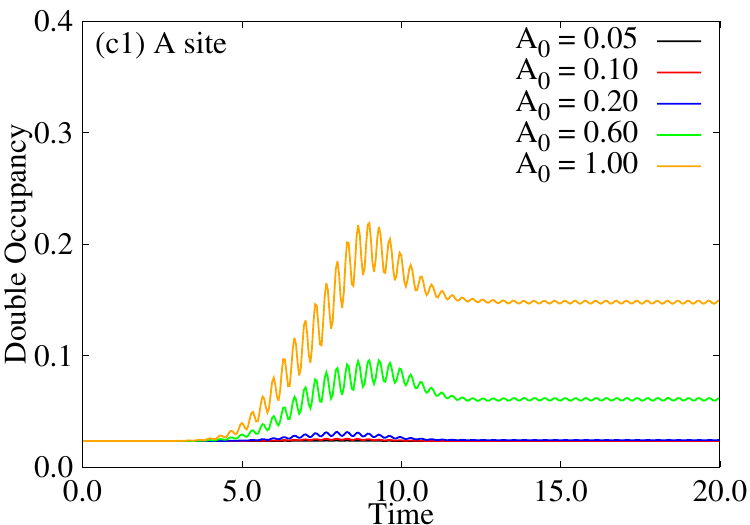}
\includegraphics[angle=-0,width=0.243\textwidth]{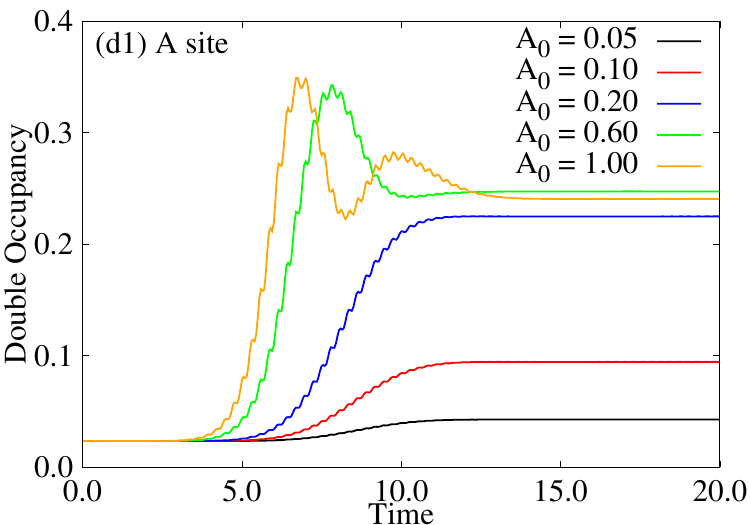}
\includegraphics[angle=-0,width=0.243\textwidth]{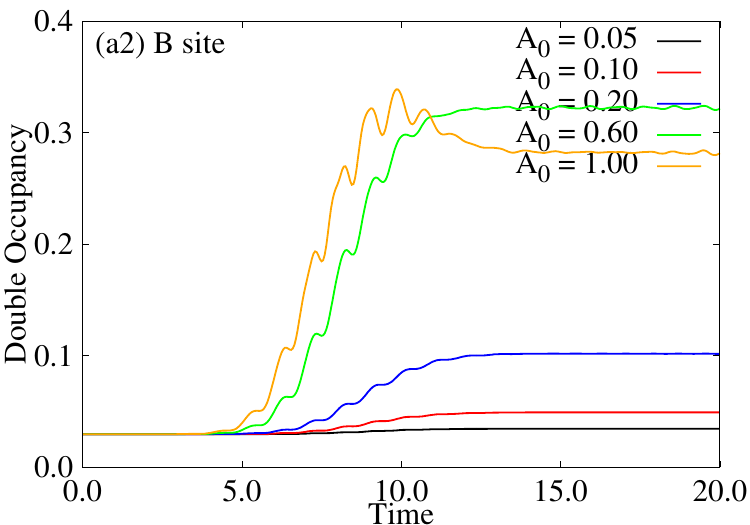}
\includegraphics[angle=-0,width=0.243\textwidth]{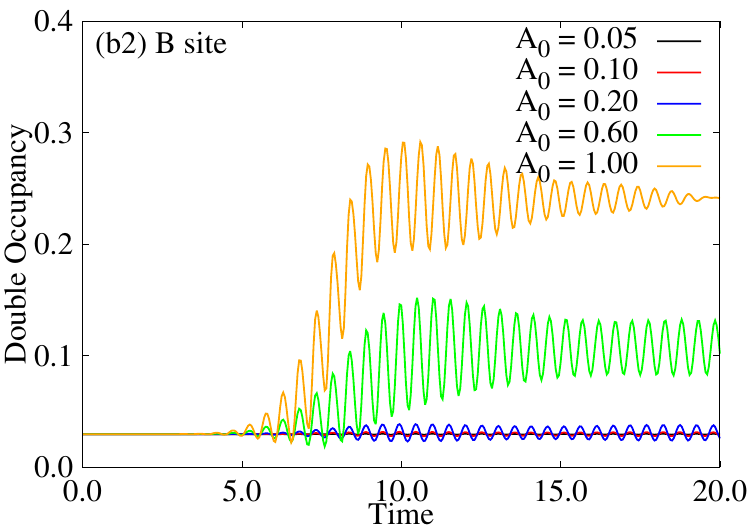}
\includegraphics[angle=-0,width=0.243\textwidth]{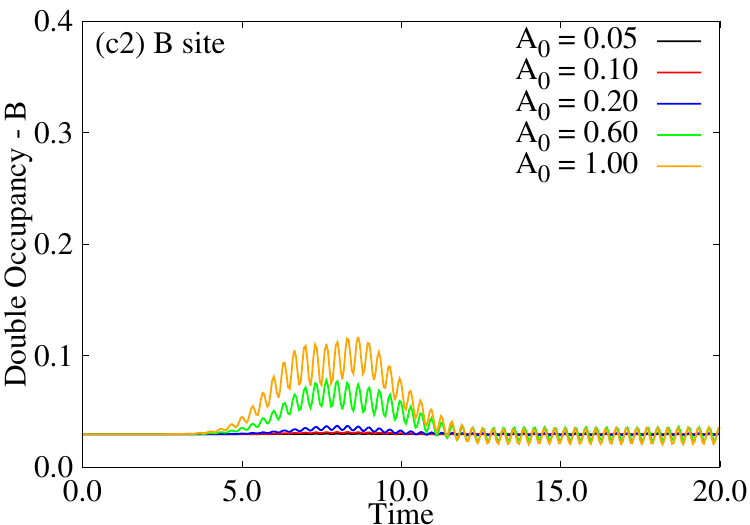}
\includegraphics[angle=-0,width=0.243\textwidth]{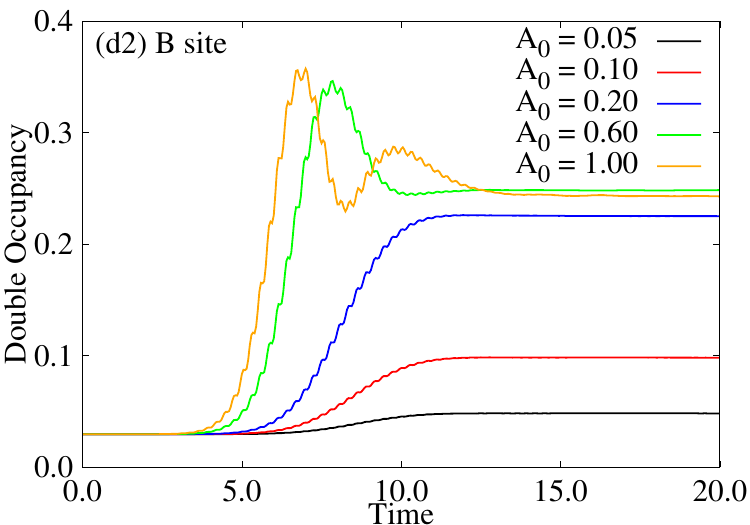}
\includegraphics[angle=-0,width=0.243\textwidth]{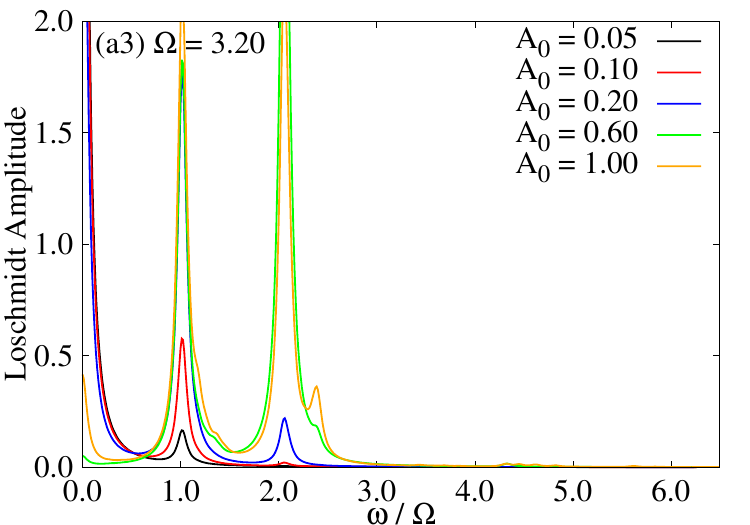}
\includegraphics[angle=-0,width=0.243\textwidth]{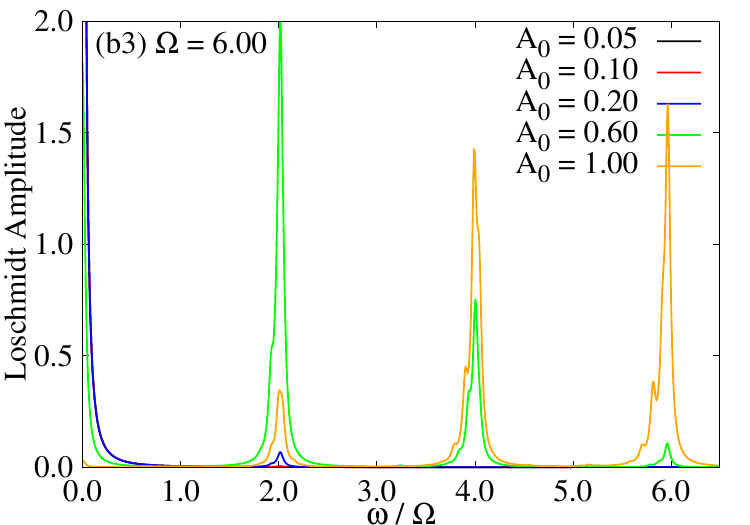}
\includegraphics[angle=-0,width=0.243\textwidth]{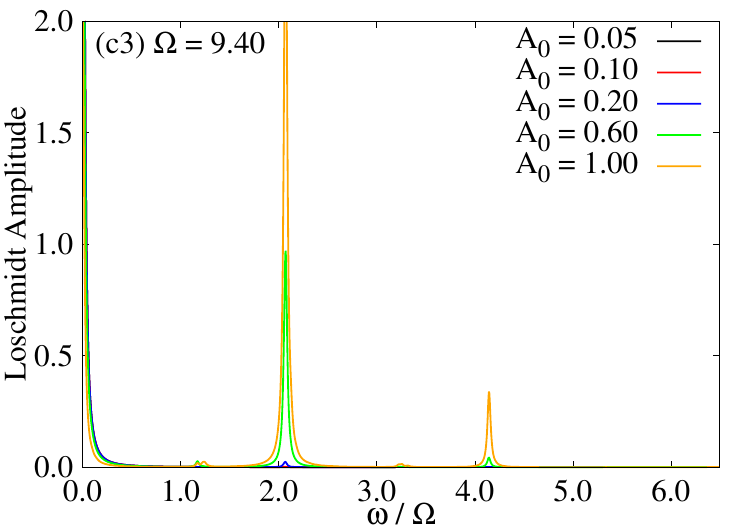}
\includegraphics[angle=-0,width=0.243\textwidth]{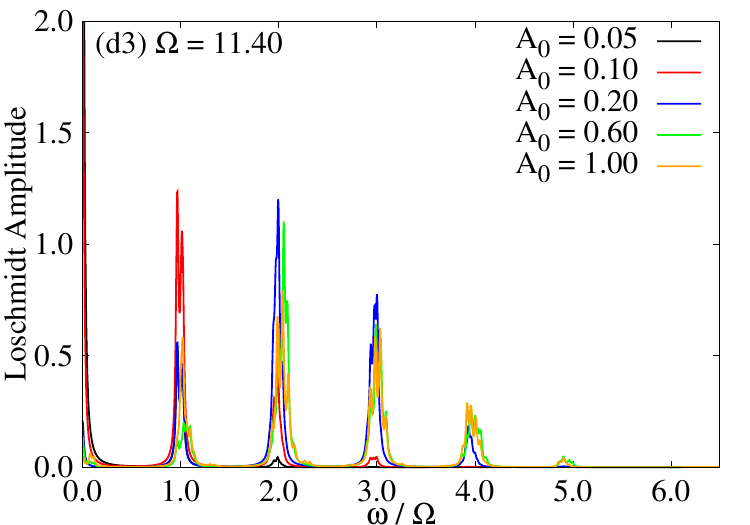}
\caption{(Color online) For the Hubbard super-lattice with fixed site-dependent Coulomb interaction strength $U_b = 3.0$ and $U_a = 18.0$, the time evolution of site dependent double occupancy with the Loschmidt amplitude in the third row for different laser amplitude $A_0$ and frequency $\Omega$. (a1-a3) $\Omega = 3.2$, (b1-b3) $\Omega = 6.0$ (c1-c3) $\Omega = 9.4$ (d1-d3) $\Omega = 11.4$, respectively.}
\label{Fig:dtL}
\end{figure*}

\section{Density of states and Linear absorption spectrum}
\label{sec:eqlabs}
In this section, we focus our attention on the Hubbard superlattice in the Mott insulating phase with finite charge gap and zero spin gap\cite{Duan:jpcm2010}, where the system parameters are $U_a = 18.0$ and $U_b = 3.0$ (measured in units of $t_h$).
In Fig.\ref{Fig:eqdos}(a), we plot the sublattice-specified density of states in Eq.\eqref{eq:eqdos} at zero temperature using the Lanczos method. 
Evidently, the particle-hole symmetries are observed for both A and B sublattices, with each site singly occupied (half-filling).
The energy gap at the Fermi energy ($E_F = 0.0$) confirms that the system is an insulator.  Note the density of states exhibits four peaks for each sublattice.
For the A-sublattice, we observe the lower and upper Hubbard band at approximately $\pm U_a/2 = \pm 9.0$ and the two hybridization bands around $\pm U_b/2 = \pm 1.5$. 
In contrast, the upper and lower Hubbard band for the B-sublattice is situated at $\pm U_b/2 = \pm 1.5$ with tiny hybridization band at $\pm U_a/2 = \pm 9.0$. 

To study effect of a laser drive on the equilibrium system (taking it out-of-equilibrium), we first study the energy absorption of the superlattice system at low laser intensity where linear response theory applies. 
In Fig.\ref{Fig:eqdos}(b), the linear absorption spectrum $\alpha(\omega)$ is calculated using Eq.\eqref{eq:labs}, where Fermi's golden rule applies\cite{Watzenbock:prb2022}. 
The positions of the peaks in the linear absorption spectrum (one-photon excited states) are $\omega \approx 3.2, 11.4$ (two peaks at $\omega \approx 11.0, 11.5$ are merged into a single peak). 
The peak at $\omega \approx 3.2$ corresponds to the energy difference between the lower Hubbard band of the 
B-sublattice and the upper hybridization peak of the A-sublattice, or inversely, from lower hybridization peak of 
A-sublattice to the upper Hubbard band of B-sublattice. 
The peak at $\omega \approx 11.4$ corresponds to the energy difference between the lower Hubbard band of 
the B-sublattice and the upper Hubbard band of the A-sublattice, or inversely, from lower Hubbard band of 
A sublattice to the upper Hubbard band of B-sublattice. 
The physical picture can be clearly understood as follows: Since the hopping between the nearest neighboring A-B sites is dominate, that process will contribute most to the optical excitation.
By contrast, the next-nearest neighbor A-A or B-B hopping, which is a second order process (compared with the nearest neighbor A-B first order hopping process) is weaker.

\section{Non-equilibrium driven behavior in the Mott insulating regime}
\label{sec:neqdynamics}
To gain an overall picture of photon absorption in the system, 
we plot the energy following the pulse ($t > 15.0$) as a function of laser frequency $\Omega$ and laser intensity $A_0$ in Fig.\ref{Fig1}(a). Note the energy remains unchanged after the laser pulse has passed since the time evolution operator is unitary (no energy is being added or removed from the system). 
A linear response behavior is observed for small laser intensity for the one-photon resonance frequencies $\Omega \approx 3.2, 11.4$, which were clarified previously by the linear absorption spectrum in Fig.\ref{Fig:eqdos}(b). 
To make the linear response regime clear, we plot the energy as a function of laser intensity with fixed laser frequency $\Omega = 3.2, 6.0, 9.4, 11.4$ in Fig.\ref{Fig1}(b). 
The total energy exhibits linear energy absorption in the regime $A_0 < 0.30$ for laser frequencies $\Omega=3.2, 11.4$. 
In contrast, the total energy does not change until $A_0  \geq 0.2$ for laser frequencies $\Omega = 6.0, 9.4$, which suggest single photon absorption is absent for a lower intensity laser and multi-photon excitation occur for higher laser intensity.  

To provide more details of the optical excitation process, we focus our attention on the non-equilibrium evolution of total energy and double occupancy of the system with the super-lattice (staggered Coulomb interaction). 
The time-dependent energy as a function of laser frequency are plotted in the top row of Fig.\ref{Fig:ewt}(a1, b1, c1, d1), where the laser intensities are set as $A_0 = 0.1, 0.2, 0.6, 1.0$, respectively. 
The time evolution of the site-dependent double occupancy are plotted in Fig.\ref{Fig:ewt} (a2-a3) for laser intensity $A_0 = 0.1$, (b2-b3) for $A_0 = 0.2$ , (c2-c3) for $A_0 = 0.6$ and (d2-d3) for $A_0 = 1.0$.

For $A_0 = 0.1$ in the linear response regime, the photon energy is absorbed at frequency $\Omega = 11.0$ and a small energy absorption is observed at frequency $\Omega = 3.2$, which are consistent with the equilibrium calculation of the linear absorption spectrum in Fig.\ref{Fig:eqdos}(b).
By inspecting the site-specific double occupancy as a function of time and laser frequency, we find that an increase of double occupancy on both site-A and site-B at frequency $\Omega = 11.4$ will contribute to the total energy absorption. While for laser frequency $\Omega = 3.2$, the energy absorption mainly comes from the enhancement of double occupancy only at site B. The behaviors above can be explained with the following physical picture: For laser frequency $\Omega = 11.4$, the single photon absorption induced optical excitation will move electrons from the lower Hubbard band of the A sublattice to the upper Hubbard band of the B sublattice, which enhances the double occupancy of the B-site.
Correspondingly, the photon can also excite electrons from the lower Hubbard band of B sub-lattice to the upper Hubbard band of A sublattice, enhancing the A-site double occupancy simultaneously.
By contrast, the physics is different for optical excitation with laser frequency $\Omega = 3.2$. The single photon process will excite electrons from the lower Hubbard band of B to the upper hybridization band of A, enhancing the kinetic energy while not affecting the double occupancy of the A-site. 
However, the optical excitation from the lower hybridization A band to the upper Hubbard bands of the B sublattice will enhance the double occupancy of the B site. 
As a result, the system will exhibit site-selective doublon-holon dynamics at $\Omega=3.2$  which results from a single photon excitation.
Further increasing the laser intensity to $A_0 = 0.2$ will not introduce new physics, except it enhances the effects observed with $A_0 = 0.1$, indicating that $A_0 = 0.2$ is still situated in the linear response region 
of optical excitation. 

Increasing the laser intensity to $A_0 = 0.6$, besides the enhancement of energy absorption at $\Omega = 3.2, 11.4$ observed before, results in energy absorption at
 at laser frequencies $\Omega = 9.4, 6.0, 1.5$ respectively. 
The extra absorption frequencies can be explained by multi-photon absorption processes, 
which indicate $A_0 = 0.6$ is out of the linear response regime.
For $\Omega = 6.0, 1.5$, the double-photon processes  correspond to the excitation of single photon process of $\Omega = 11.4, 3.2$, respectively (as the frequencies are roughly doubled).
However, the two photon process for $\Omega = 9.4$ is different from the other two frequencies. 
For $\Omega = 9.4$, an apparent enhancement of double occupancy is observed for the A-site while not for the B-site. 
The short time process is due to an excitation from the lower Hubbard band of A sublattice to the upper Hubbard band of the B sublattice, which enhances the double occupancy of B site.  At subsequent times, the quasi-particle is further excite to the upper Hubbard of the A sublattice, which results in a reduction of the double occupancy of the B site and an increment of the A site.
Thus, the double occupancy of the A sites is enhanced while B sites do not change in long time regime (which is induced by the two-photon process in the relative long time regime). 
Further increasing the laser intensity up to $A_0 = 1.0$ will enhance the observed doublon-holon dynamics of $A_0=0.6$, and a triple photon absorption is observed at $\Omega = 6.5$ (confirmed with Loschmidt amplitude, not shown).
In summary, a clear frequency selection behavior is observed from the energy absorption and site-specific double occupancy enhancement, while the quasi-particle physical picture for the selective dynamics can be different for different pump frequencies. 

To make the laser frequency selection in the dynamical behavior more clear, the site-specific double occupation as a function of time with different laser intensities are plotted in Fig.\ref{Fig:dtL}, with the Loschmidt amplitude (see also Sec.\ref{sec:gla}) as a function of $\omega$ plotted in the third row. 
For laser frequency $\Omega = 3.2$, the double occupancy of the A-site remains nearly unchanged while the B-site is significantly enhanced with increased laser intensity. For laser frequency $\Omega = 8.0$ (not shown), the double occupancy of both the A- and the B-site are unchanged with light. The physics can be explained using the quasiparticle picture of photonic excitation in Ref. \cite{Maislinger:prb2022}.
For laser frequency $\Omega = 11.40$, the double occupancy of both A-sites and B-sites are strongly enhanced.
For laser frequency $\Omega = 6.0$, we find that the double occupancy does not change for small laser intensity $A_0=0.05, 0.10, 0.20$, while for $A_0 = 0.60, 1.00$, the double occupancy of both A and B-site increase in a similar fashion, which can be explained as multiphoton processes. The multi-photon processes can be identified as double-photon processes by looking at the Loschmidt amplitude, peaking at $2\Omega, 4\Omega, 6\Omega$.
 Further, the oscillation frequency is independent of the laser intensity, which is different from the Rabi-like behavior (oscillation frequency increase with laser intensity) observed in the pumped two-dimensional Hubbard cluster\cite{Okamoto:sr2021}.
 At the laser frequency $\Omega = 9.4$, the characterized two-photon process is a 
 one-photon process from the lower Hubbard band of A to the upper Hubbard band of B, with a subsequent excitation to upper Hubbard band of A. 
\begin{figure*}[ht]
\centering
\includegraphics[angle=-0,width=0.245\textwidth]{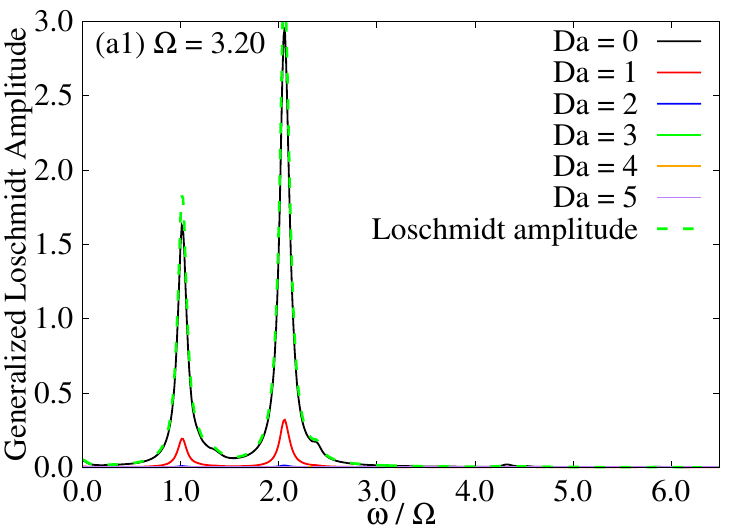}
\includegraphics[angle=-0,width=0.245\textwidth]{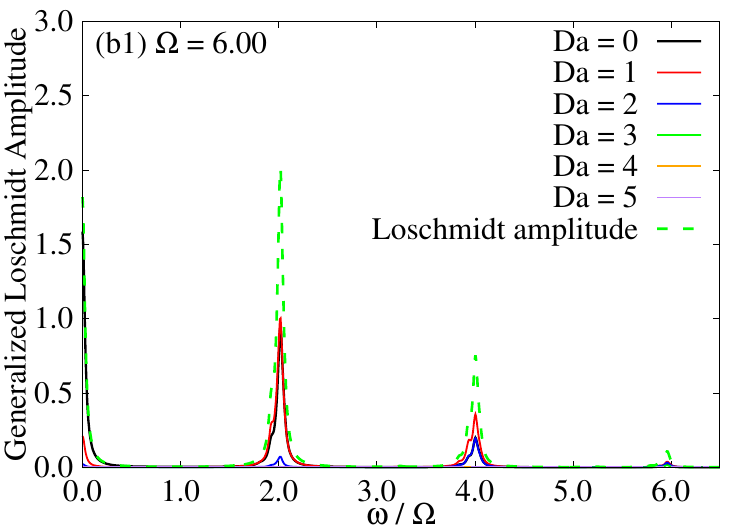}
\includegraphics[angle=-0,width=0.245\textwidth]{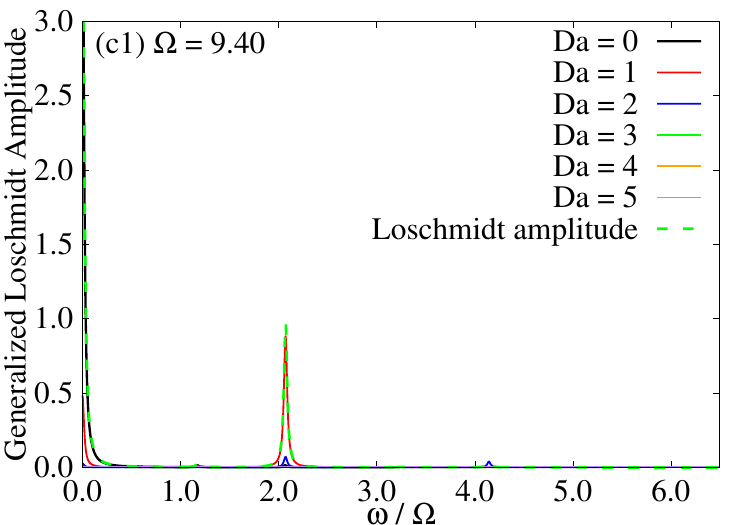}
\includegraphics[angle=-0,width=0.245\textwidth]{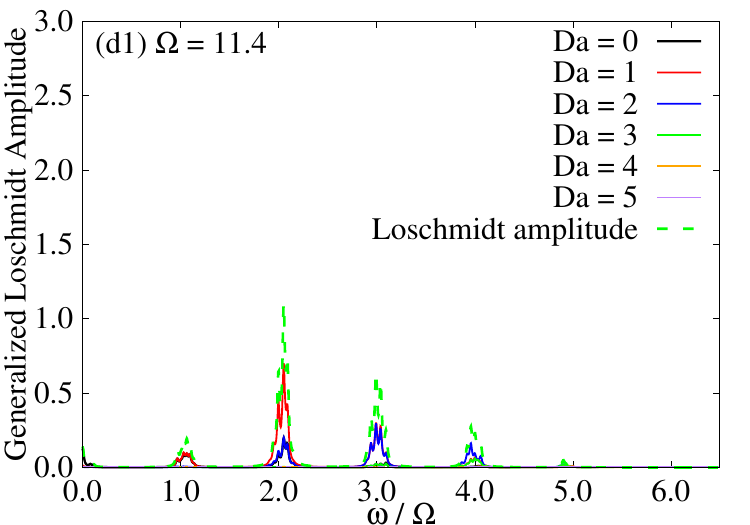}
\includegraphics[angle=-0,width=0.245\textwidth]{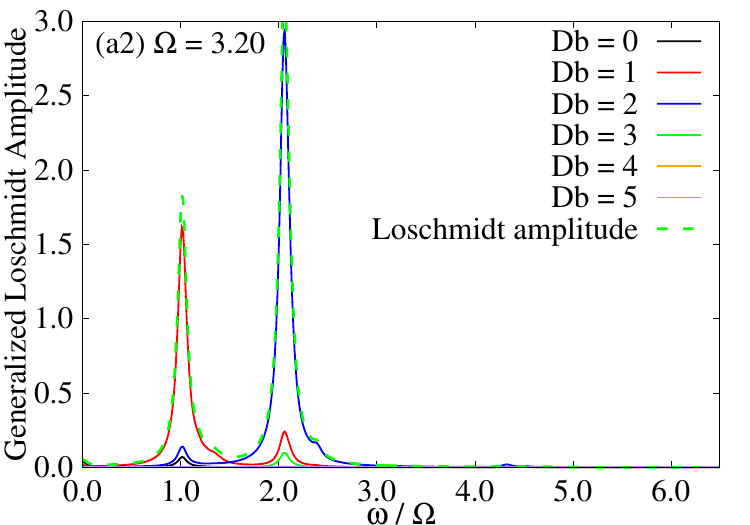}
\includegraphics[angle=-0,width=0.245\textwidth]{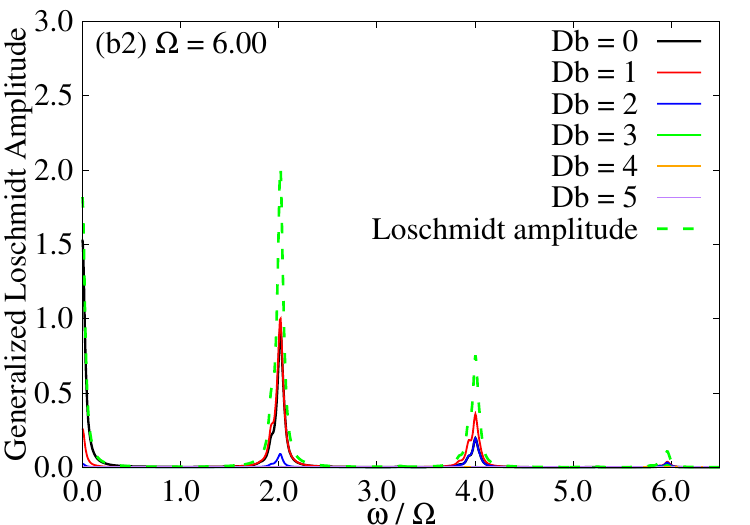}
\includegraphics[angle=-0,width=0.245\textwidth]{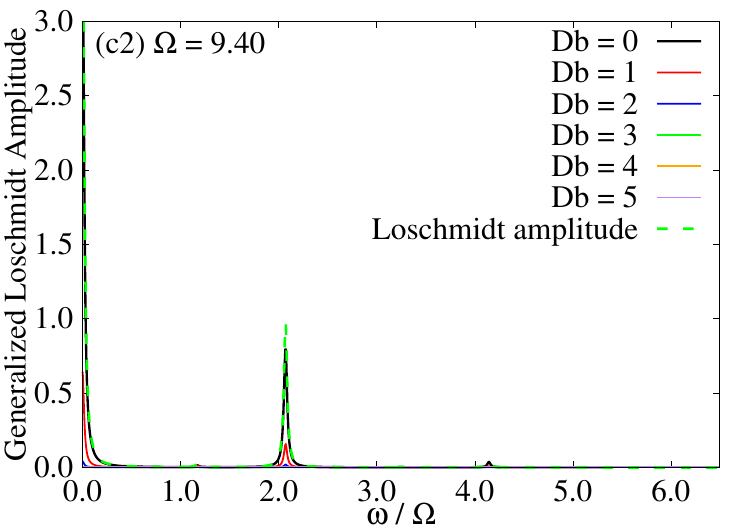}
\includegraphics[angle=-0,width=0.245\textwidth]{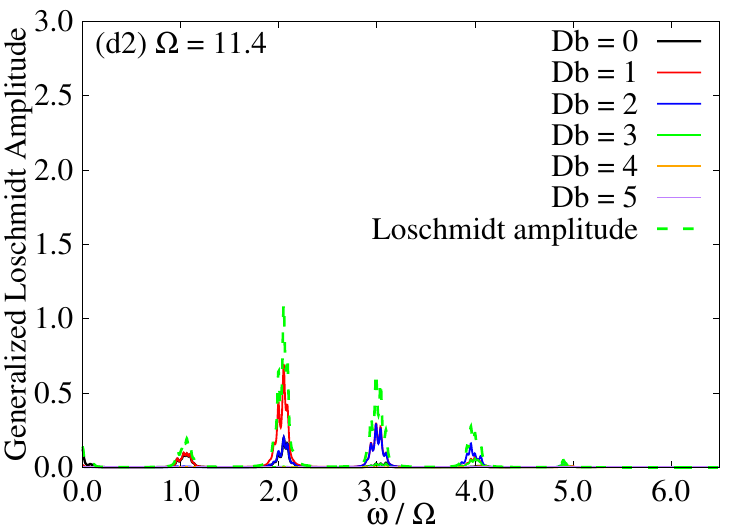}
\caption{(Color online) For the Hubbard super-lattice with fixed site-dependent Coulomb interaction strength $U_b = 3.0$ and $U_a = 18.0$, the generalized Loschmidt amplitude as a function of energy for different laser frequency, (a1-a2) $\Omega = 3.2$, (b1-b2) $\Omega = 6.0$ (c1-c2) $\Omega = 9.4$ (d1-d2) $\Omega = 11.4$, respectively. The green dash line is the Loschmidt amplitude, which is a guidance of sum rule in Eq.\eqref{eq:Lsum} for eye.} 
\label{Fig:Lwt}
\end{figure*}

\section{The Generalized Loschmidt Amplitude}
\label{sec:gla}
In the context of site selective doublon-holon dynamics, it is interesting to know which energies will contribute significantly to the double occupancy enhancement of the A-site or the B-site after the laser pulse has passed ($t=40.0$ used here).
To this end, we followed previous work\cite{Watzenbock:prb2022} and defined the generalized Loschmidt amplitude with respect to energy and site-specified double occupancy,
\begin{align}
    L_{\hat{H}\!\hat{D}_\alpha} (\omega, D, t) = & \sum_{m,n} \langle\psi(t)|E_n\rangle\langle E_n|D_m\rangle\langle D_m|\psi(t)\rangle \nonumber\\
    &\times \delta(\omega - E_n) \delta(D - D_m^\alpha).
    \label{eq:gla}
\end{align}
where $\hat{D}_\alpha = \sum_{i\in \alpha} \hat{n}_{i\uparrow} \hat{n}_{i\downarrow}$ ($\alpha = a, b$ for A, B sub-lattice) and $\hat{D}_\alpha |D_m\rangle = D_m^\alpha|D_m\rangle$ with $|D_m\rangle$ the many-body Fock basis with double occupancy $D_m = D_m^a + D_m^b$.  Here $|E_n\rangle$ are energy eigenstates of the Hamiltonian with eigenvalue $E_n$. Due to the fact that the eigenspectrum and eigenstates of double occupancy are \textit{a priori} known, a computational cheaper way to obtain the same information as Eq.\eqref{eq:gla} is using projectors\cite{Watzenbock:prb2022},
\begin{align}
    L_{H} (\omega, \hat{P}_{m}) = \sum_{n} \langle\psi(t)|E_n\rangle\langle E_n|\hat{P}_m|\psi(t)\rangle \delta(\omega - E_n) 
\end{align}
where $\hat{P}_m = |D_m\rangle \langle D_m|$ is the projector operator onto states with double occupancy, $D_m \in \{0, 1, 2, \cdots, N/2\}$,
\begin{align}
    \sum_{D_m=0}^{N/2}  L_{H} (\omega, \hat{P}_{m}) =  L(\omega).
    \label{eq:Lsum}
\end{align}
The total double occupancy is expressed as,
\begin{align}
    \sum_{D_m=0}^{N/2} D_m \int d\omega L_{H}(\omega, \hat{P}_{m}) =  \langle \hat{D} \rangle.
\end{align}
The site-specific double occupancy is expressed as,
\begin{align}
    \sum_{D_m=0}^{N/2} D_m^\alpha \int d\omega L_{H}(\omega, \hat{P}_{m}) =  \langle \hat{D}_\alpha \rangle,
\end{align}
where $\alpha=a,b$.

In Fig.\ref{Fig:Lwt}, we plot the generalized Loschmidt amplitude as a function laser frequency, where the laser amplitude is fixed at $A_0 = 0.60$. For laser frequency $\Omega = 3.2$, the single and double photon processes will induce a hopping within the $D_a = 0$ subspace, while the single photon process will include the state with one double occupancy in the B sublattice, and two photon processes will induce two double occupancies on the B sites. As a result, the A site double occupancy do not change and the B site double occupancy is enhanced. 

For laser frequency $\Omega = 6.0$, the single photon processes are absent and only double-photon process will enhance the $D_a = 1$ subspace. Note that the four photon process is a subsequent two-photon process (one following the other). 
For laser frequency $\Omega = 9.4$, the single photon processes are absent and only a double-photon process will enhance the $D_b = 1$ subspace, and the the $D_a = 0$ subspace. Note that the four photon process is a subsequent two-photon process (one following the other).
For laser frequency $\Omega = 11.4$, the single photon processes are observed and a single-photon will enhance the $D_b = 1$ subspace, and the $D_a = 1$ subspace. Note that the four photon process is a subsequent two-photon process (one following the other).

\section{Conclusion and Discussion}
\label{sec:concl}
In this paper, we study the doubon-holon dynamics in the Hubbard superlattice with alternatively spatially modulated (staggered) on-site Coulomb interaction driven by a laser pulse.
The Coulomb interaction parameters are set as $U_a = 18.0$ and $U_b = 3.0$ for odd and even sites of the one-dimensional chain with periodic boundary conditions adopted. 
In equilibrium, the site-specific density of states is studied and shown to exhibit four peaks with lower and upper Hubbard bands and two hybridization peaks. Within linear response theory, we find two main resonance frequency--$\Omega = 3.2$ and $\Omega = 11.4$--by studying the linear absorption spectrum. 
Focusing on the dynamics of site-specific double occupancy, we study the time evolution of double occupancy at the A and B sites, in the plane of laser intensity and frequency. For small laser intensity, the system is in the linear response theory regime, where the resonance frequencies are $\Omega = 3.2$ and $\Omega = 11.4$. At $\Omega = 3.2$, the double occupancy of the A-sites remain approximately unchanged, while the double occupancy of the B-sites is largely enhanced. 
The site-selective doublon-holon dynamics observed above is due to the fact that the upper hybridization of the A band is mainly singly occupied, while the upper Hubbard band of B is mainly doubly occupied. 
At $\Omega = 11.4$, the double occupancy of both A-sites and B-sites are increased simultaneously by the laser, which are explained as the resonance excitation from the lower Hubbard of A(B) to upper the Hubbard band of B(A).
For higher laser intensity beyond the linear response regime, multi-photon processes will play an important role in the doublon-holon dynamics. 
At $\Omega = 6.0$, the double occupancy of the A- and B-sites remain almost unchanged (oscillating around their equilibrium value at time $t=0.0$) for laser intensity $A_0 = 0.05, 0.10, 0.20$, while for laser intensity $A_0 = 0.6, 1.0$, the double occupancy of both A, B-sites increase simultaneously. 
We conclude the double photon processes at $\Omega = 6.0$ for high laser intensity will induced hopping events observed for single photon processes around $\Omega = 11.4$. 
Furthermore, at $\Omega = 9.4$, the double occupancy of the A-sites are increased significantly while that of the B-sites remain almost unchanged, which is what is observed in site-selective doublon-holon dynamics. The physical picture for $\Omega = 9.4$ is different from the one observed at $\Omega = 3.2$, where one hopping event is introduced to explain the dynamics. The phenomena for $\Omega = 3.2$ is explained with two subsequent hopping events: the hopping from the lower Hubbard band of A to the upper Hubbard band of B, followed by a second hopping from the upper Hubbard band of B to the upper Hubbard band of A.
The physical pictures described above have been confirmed by direct evaluation of the Loschmidt amplitudes. Our theoretical study suggests strategies to engineer the doublon-holon dynamics with specified laser parameters.

\appendix
\setcounter{equation}{0}
\renewcommand\theequation{A\arabic{equation}}
\setcounter{figure}{0}
\renewcommand\thefigure{A\arabic{figure}}
\section{Particle-Hole symmetry of the Hubbard superlattice}
Introducing the particle-hole transformation in a bipartite lattice (A-B sublattice),
\begin{align}
    c_{i\sigma}^\dagger \rightarrow \eta_i d_{i\sigma}^{}, \quad c_{i\sigma}^{} \rightarrow \eta_i d_{i\sigma}^\dagger,
\end{align}
where
\begin{align}
    \eta_i = \left\{ \begin{array}{rl}
             1 & \mathrm{if}\ i \in \mathrm{A}, \\
            -1 & \mathrm{if}\ i \in \mathrm{B}.
    \end{array}\right.
\end{align}
Consider the local terms in one unit cell,
\begin{align}
    H_{\mathrm{loc}} &=  U_a  n_{a\uparrow}n_{a\downarrow} +  U_b n_{b\uparrow}n_{b\downarrow}\nonumber\\
                     &+ \epsilon(n_a - n_b) - \mu (n_a + n_b).
\label{eqH}
\end{align}
The particle-hole transformed Hamiltonian is,
\begin{align}
    \tilde{H}_{\mathrm{loc}} &= U_a  (1-n_{a\uparrow})(1-n_{a\downarrow}) +  U_b (1-n_{b\uparrow})(1-n_{b\downarrow})\nonumber\\
             &+\epsilon(n_b - n_a) - \mu (4- n_a - n_b).
\label{eqHtilde}
\end{align}
The difference between $H_{\mathrm{loc}}$ and its particle-hole transformation $\tilde{H}_{\mathrm{loc}}$ is,
\begin{align}
    H_{\mathrm{loc}} - \tilde{H}_{\mathrm{loc}}
             &= (U_a + 2\epsilon - 2\mu)  n_a + (U_b - 2\epsilon - 2\mu)  n_b \nonumber\\
             &+ (4\mu - U_a - U_b).
\label{eqHtilde}
\end{align}
For the particle-hole symmetry, we have,
\begin{align}
      \epsilon = (U_a - U_b) / 4, \quad \mu = (U_a + U_b) / 4.
\label{eqHtilde}
\end{align}
\setcounter{equation}{0}
\renewcommand\theequation{B\arabic{equation}}
\section{Position of lower and upper Hubbard band of the Hubbard superlattice}
In the homogeneous one band Hubbard model at half-filling, the position of lower and upper Hubbard bands are situated at $-U/2$ and $U/2$ in equilibrium, which can be understood by calculating the lesser Green's function and greater Green's function at strong Coulomb interaction limit.
The lesser Green's function is defined as,
\begin{align}
    G_\uparrow^<(\omega) = -\sum_n \frac{\langle \Psi_g| c_\uparrow^\dagger | n\rangle\langle n|c_\uparrow |\Psi_g\rangle}{\omega + i\eta - (E_0 - E_n)},
\end{align}
where $E_0$ and $|\Psi_g\rangle = |\uparrow, \downarrow, \uparrow,\downarrow,\cdots\rangle$ are the ground state energy and vector in the subspace $(N_\uparrow, N_\downarrow)$, $E_n$ and $|n\rangle$ are the eigen-energy and eigen-vector in the subspace $(N_\uparrow-1, N_\downarrow)$. As a result, the $E_0 - E_n$ constitute the position of lower Hubbard band.
The greater Green's function is defined as,
\begin{align}
    G_\uparrow^>(\omega) = +\sum_m \frac{\langle \Psi_g| c_\uparrow |m\rangle\langle m|c_\uparrow^\dagger |\Psi_g\rangle}{\omega + i\eta - (E_m - E_0)},
\end{align}
where $E_0$ and $|\Psi_g\rangle$ are the ground state energy and vector in the subspace $(N_\uparrow, N_\downarrow)$, $E_m$ and $|m\rangle$ are the eigenenergy and eigenvector in the subspace $(N_\uparrow+1, N_\downarrow)$. As a result, the energy difference $E_m - E_0$ constitutes the position of the upper Hubbard band with one more doubly occupied site.

For the half-filled Hubbard superlattice with staggered Coulomb interactions, its atomic limit Hamiltonian in one unit cell is written as,
\begin{align}
    H_\mathrm{loc} &= U_a \left(n_{a\uparrow} - 1/2\right)\left( n_{a\downarrow} - 1/2\right) \nonumber\\
    &+ U_b \left(n_{b\uparrow} - 1/2\right)\left( n_{b\downarrow} - 1/2\right).
\end{align}
The ground state $|\uparrow_a,\downarrow_b\rangle$ ($|\downarrow_a,\uparrow_b\rangle$) energy in $(N_\uparrow = 1, N_\downarrow = 1)$ subspace is $-(U_a+U_b)/4$. 
For the A-sublattice (where the newly added or deleted electron is from the A site), in subspace $(N_\uparrow = 0, N_\downarrow = 1)$, the ground state energy is $(U_a-U_b)/4$, which conclude the lower Hubbard band is situated at $-U_a/2$. In subspace $(N_\uparrow = 2, N_\downarrow = 1)$, the ground state energy is $(U_a-U_b)/4$, where the upper Hubbard band is situated at $U_a/2$. 
For the B-sublattice (where the newly added or deleted electron is from the B-site), at subspace $(N_\uparrow = 1, N_\downarrow = 0)$, the ground state energy is $(U_b-U_a)/4$, where the lower Hubbard band is situated at $-U_b/2$. In subspace $(N_\uparrow = 1, N_\downarrow = 2)$, the ground state energy is $(U_b-U_a)/4$, where the upper Hubbard band is situated at $U_b/2$. 
\begin{figure*}[ht]
\centering
\includegraphics[angle=-0,width=0.24\textwidth]{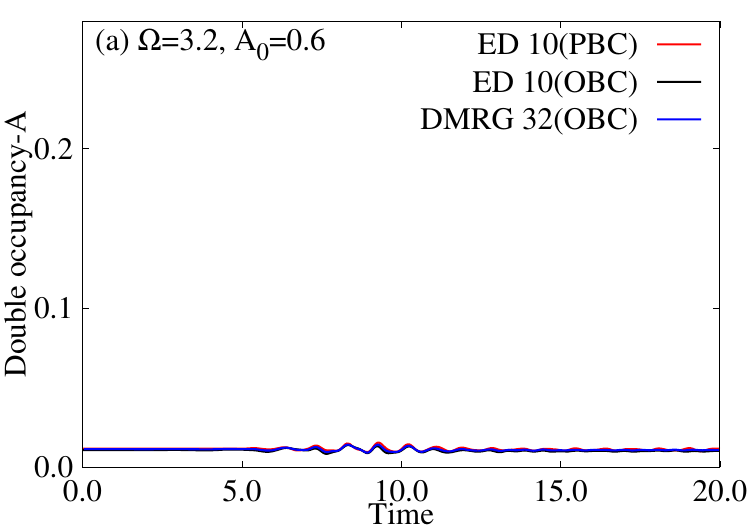}
\includegraphics[angle=-0,width=0.24\textwidth]{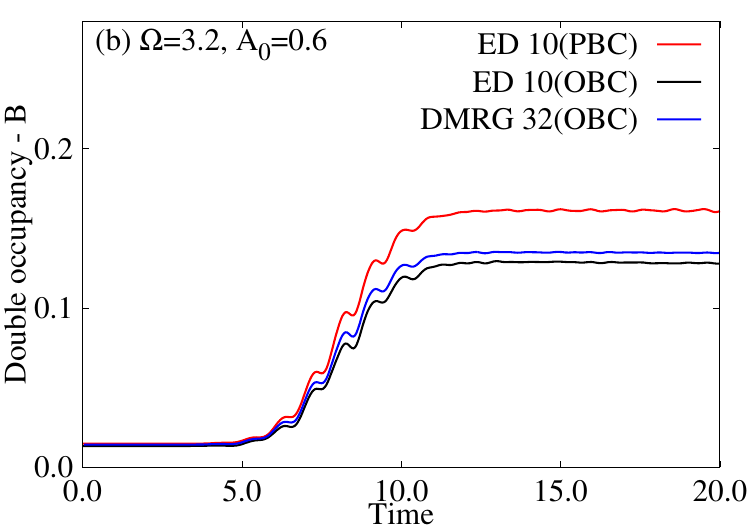}
\includegraphics[angle=-0,width=0.24\textwidth]{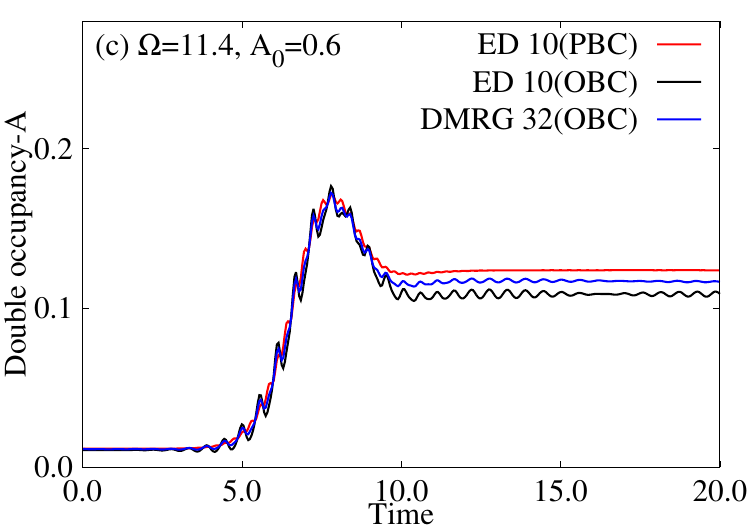}
\includegraphics[angle=-0,width=0.24\textwidth]{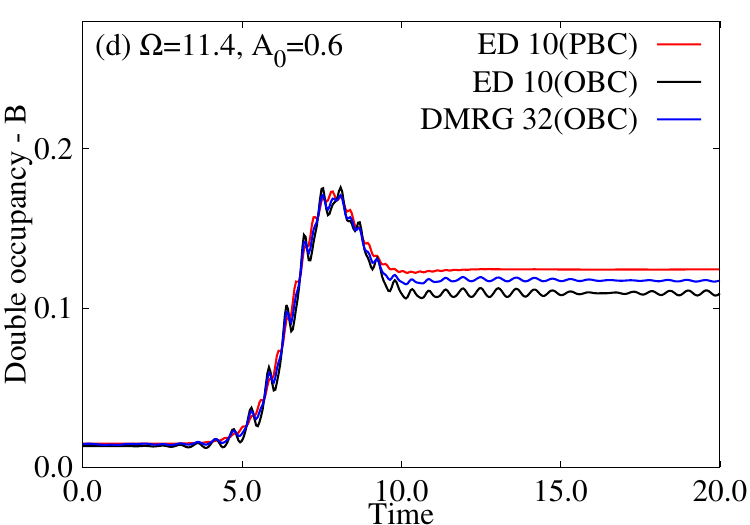}
\caption{(Color online) For the Hubbard super-lattice with fixed site-dependent Coulomb interaction strength $U_b = 3.0$ and $U_a = 18.0$, the time evolution of site dependent double occupancy. (a-b) $\Omega = 3.2$ A-site and B-site, (c-d) $\Omega = 11.0$ A-site and B-site, respectively.}
\label{Fig:itensor}
\end{figure*}
\section{Finite size effects}
To determine the finite size effects in the system, we compare our numerical results of the pumped one-dimensional chain with 32 sites, solved with density matrix renormalization group (DMRG) theory using Itensor\cite{itensor,YangM:prb2020}. 
The system parameters are $U_a=18.0, U_b=3.0$. The laser frequency is set as $\Omega=3.2$ in Fig.\ref{Fig:itensor} (a-b) and $\Omega=11.0$ in Fig.\ref{Fig:itensor} (c-d).
The double occupancy calculated with DMRG and exact diagonalization show good agreement with each other, which indicate the finite size effects in this work are not severe.

\acknowledgements
We acknowledge helpful discussions with Tao Li, Huan Li, Xiaojun Zheng. We gratefully acknowledge funding from the Natural Science Foundation of Guangxi Province Grant No. 2020GXNSFAA297083, GuiKe AD20297045 and the National Natural Science Foundation of China (Grant No. 11904143, No. 12174168, No. 12047501, No. 12147211, and No. 12065002), G.A.F. gratefully acknowledges funding from US National Science Foundation grant DMR-2114825 and the Alexander von Humboldt Foundation.

\bibliography{uaubref}
\end{document}